\documentclass[11pt]{article}
\usepackage{graphicx}
\begin{document}

\title{Possible Molecular Structure of the Newly Observed $Y(4260)$} \maketitle

\centerline{Xiang Liu$^1$, Xiao-Qiang Zeng$^1$, and Xue-Qian
Li$^1$}

\vspace{0.8cm}

1. Department of Physics, Nankai University, Tianjin 300071, China

\begin{abstract}
We suggest that the newly observed resonance $Y(4260)$ is a
$\chi_{c}-\rho^0$ molecule, which is an isovector. In this
picture, we can easily interpret why $Y(4260)\rightarrow
\pi^+\pi^-J/\psi$ has a larger rate than $Y(4260)\rightarrow D\bar
D$ which has not been observed, and we also predict existence of
the other two components of the isotriplet and another two
possible partner states which may be observed in the future
experiments. A direct consequence of this structure is that for
this molecular structure $Y(4260)\rightarrow \pi^+\pi^-J/\psi$
mode is more favorable than $Y(4260)\rightarrow K\bar KJ/\psi$
which may have a larger fraction if other proposed structures
prevail.
\end{abstract}

\section{General discussion}

The BaBar collaboration has recently announced that a very
intriguing new state/structure $Y(4260)\rightarrow
\pi^+\pi^-J/\psi$ is observed in $e^+e^- \rightarrow ISR
\pi^+\pi^-J/\psi$, where ISR stands for Initial-State Radiation
\cite{Babar}. Their results indicate that $Y(4260)$ has
spin-parity $J^{PC}=1^{--}$. Its mass and width are
\begin{eqnarray*}
m=4.26\;\mathrm{GeV}/\mathrm{c}^{2}\;\;\;\mbox{and}\; \;\;
\Gamma\sim 90\;\mathrm{MeV}/\mathrm{c}^2.
\end{eqnarray*}
An enhancement near 4.26 $GeV/c^2$ is clearly observed in $e^+e^-
\rightarrow ISR \pi^+\pi^-J/\psi$ channel, but has not been
observed in $e^+e^-\rightarrow hadrons$, especially not in the
$D_{(s)}\bar D_{(s)}$ channel. It may imply that the branching
ratio of $Y(4260)\rightarrow J/\psi\pi^{+}\pi^{-}$ is much larger
than that of $Y(4260)\rightarrow D\bar D$\cite{Zhu}.

Since its remarkable characteristics, this discovery stimulates
intensive discussions about the structure of $Y(4260)$, especially
if there is some new physics involved.

First, it exists in the energy-range of $\psi$ family, and one may
expect that it involves both $c$ and $\bar c$ since in its
strong-decay products there is no single charm(anti-charm). On
other aspects, $Y(4260)\rightarrow J/\psi\pi^{+}\pi^{-}$ is a
three-body decay whereas $Y(4260)\rightarrow D\bar D$ is a
two-body decay, and usually the former is about two orders smaller
than the later due to a suppression from the phase space of final
state. However, the data indicate a reversed pattern. This
characteristic challenges our theory and demands a plausible
interpretation.

The newly observed resonance is very unlikely to be accommodated
in the regular $c\bar c$ structure even with higher radial and/or
orbital excitations. It may be a clear signal for a new structure.
In Ref. \cite{Zhu,Kou,Close}, the authors analyze the
characteristics of $Y(4260)$ and proposes that $Y(4260)$ is
perhaps a hybrid charmonium. Different from this explanation,
Maiani et al. consider that the new resonance $Y(4260)$ may be the
first orbital excitation of a diquark-antidiquark state
$[cs][\bar{c}\bar{s}]$ \cite{Maiani}. With a different point of
view, instead of supposing $Y(4260)$ to be an exotic state,
Llanes-Estrada \cite{Felipe} proposes that the experimental
evidence is not compelling to declare this state an exotic, and
can be fitted within a standard quarkonium scenario.

In this work, we propose an alternative possibility that $Y(4260)$
is an s-wave molecular state of $\rho-\chi_{c1}(1P)$, which is an
isovector. In this framework, we can naturally explain why the
branching ratio of $Y(4260)\rightarrow J/\psi\pi^{+}\pi^{-}$ is
larger than that of $Y(4260)\rightarrow D\bar{D}$. Meanwhile, we
further predict existence of possible partner resonances of
$Y(4260)$.

In next section we present our picture in detail and then we will
draw our conclusion and make a discussion in the last section.

\section{The molecular structure for $Y(4260)$.}

Actually, there has been a long history about the molecular
structure of hadrons. To explain some phenomena which are hard to
find natural interpretations in the regular valence quark
structure, people have tried to look for new structures beyond it.
The molecular structure is one of the possible candidates. Okun
and Voloshin studied the interaction between charmed mesons and
proposed possibilities of the molecular states involving charmed
quarks \cite{Okun}. Rujula, Geogi and Glashow  suggested that
$\psi(4040)$ is a $D^{*}\bar{D}^{*}$ molecular state
\cite{Rujula}. Moreover, the measured resonances $f_{0}(980)$,
$a_{0}(980)$ may be reasonably interpreted as $K\bar{K}$ molecules
\cite{Isgur}. It seems that at the energy region of charm,
molecular structure might be more favorable than at other energy
regions.

Therefore, before invoking some fancy structures, let us study
possibility to construct a molecular state for $Y(4260)$ and see
if it coincides with the observed characteristics.

From the Data-book \cite{PDG}, we find that three particles
$\chi_{c0}$, $\chi_{c1}$ and $\chi_{c2}$ in the $c\bar c$ meson
spectrum may be candidates for the constituents in $Y(4260)$. The
quantum numbers of $\chi_{c0}$, $\chi_{c1}$ and $\chi_{c2}$ are
$J^{PC}=0^{++},\; 1^{++}$ and $2^{++}$ respectively. If combining
them with $\rho$ meson to construct $\chi_{c}-\rho$ systems, one
can obtain states with spin-parity $1^{--}$. Meanwhile, the masses
of $\chi_{c0}$, $\chi_{c1}$ and $\chi_{c2}$ are well measured as
$3415.19\pm 0.34$ MeV, $3510.59\pm 0.10$ MeV and $3556.26\pm 0.11$
MeV, thus we have that $M_{\chi_{c}}+M_{\rho}$ is 4185 MeV, 4280
MeV and 4326 MeV respectively for $\chi_{c0}$, $\chi_{c1}$ and
$\chi_{c2}$.

For an s-wave molecular state, one should expect that the sum of
the constituent masses is closer to the mass of the resonance. The
difference is due to the interaction between the constituents
which in general results in a negative binding energy for s-wave.
For the three-combinations $\rho-\chi_{c0}$, $\rho-\chi_{c1}$ and
$\rho-\chi_{c2}$, one can observe that the mass sum of $\rho$ and
$\chi_{c1}$ is mostly close to the mass of $Y(4260)$. Based on the
above considerations, we propose that $Y(4260)$ may be a molecular
state of $\rho$ and $\chi_{c1}$. Namely the mass sum of $\rho$ and
$\chi_{c1}$ is about 20 MeV above 4260 MeV and the difference is
paid to the negative binding energy.

The decay pattern of $Y(4260)$ is the most important issue to
concern, because it may provide us the information about the
structure of $Y(4260)$.

In the Fig.1, we present the quark diagrams for
$Y(4260)\rightarrow J/\psi\pi^{+}\pi^{-}$ and $Y(4260)\rightarrow
D\bar{D}$.

\begin{figure}[htb]
\begin{center}
\begin{tabular}{ccc}
\scalebox{0.7}{\includegraphics{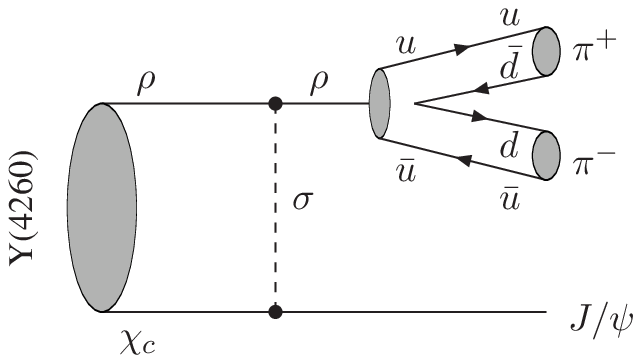}}&\hspace{0.5cm}
&\scalebox{0.7}{\includegraphics{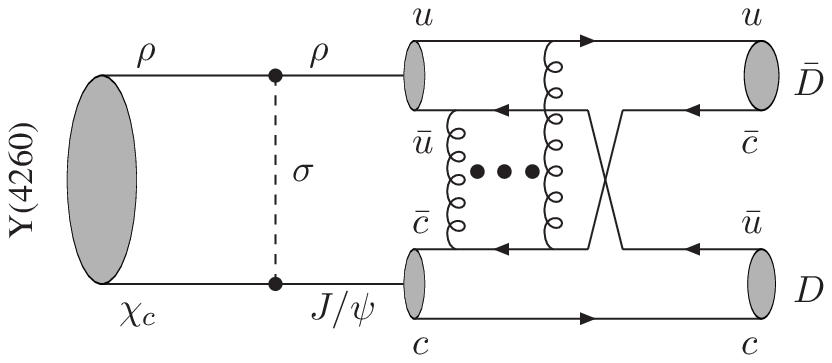}}\\
(a)&\hspace{0.5cm}&(b)
\end{tabular}
\end{center}\label{de}
\caption{(a) and (b) correspond to $Y(4260)\rightarrow
J/\psi\pi^{+}\pi^{-}$  and the $Y(4260)\rightarrow D\bar{D}$
decays respectively.}
\end{figure}

For Fig.1 (a) and (b), the transition matrix elements can be
expressed as
\begin{eqnarray}
\mathcal{M}(Y(4260)\rightarrow
J/\psi\pi^{+}\pi^{-})&=&\langle\rho^0,J/\psi|\mathcal{H}_{dis}|Y(4260)\rangle\times
\langle\pi^+\pi^-|\mathcal{H}|\rho^0\rangle,\\
\mathcal{M}(Y(4260)\rightarrow D\bar{D})&=&\langle D\bar{D}|
\mathcal{H}_{cross}|Y(4260)\rangle
\end{eqnarray}
where $\mathcal{H}_{dis}$ corresponds to the hamiltonian which
breaks the bound state $Y(4260)$ into free $J/\psi$ and $\rho^0$
via exchanging  $\sigma$ meson (maybe, exchanges of
multi-soft-gluons and even glueball of $0^{++}$ can also
contribute, but definitely $\sigma-$exchange plays the leading
role), $\mathcal{H}$ is a strong interaction which causes $\rho^0$
decay into $\pi^+\pi^-$. $\mathcal{H}_{cross}$ is an interaction,
by which quarks (antiquarks) in $\chi_c$ and $\rho^0$ exchange and
turn into hadronic $D$ and $\bar D$, in the process quark lines
cross with each other (see Fig.1 (b)).

$\chi_{c1}$ is a $1^{++}$ axial vector, $J/\psi$ is a $1^{--}$
vector and both of them are isosinglet, the couplings of
$\chi_{c1}-\pi(\rho)-\chi_{c1}$ and $\chi_{c1}-\pi(\rho)-J/\psi$
are forbidden by the isospin conservation, and only $\sigma$ of
$0^{++}$ can be exchanged and is the main contribution to the
potential which holds the constituents in a molecule. The
interactions of $\chi_{c1}-\sigma-\chi_{c1}$ and
$\chi_{c1}-\sigma-J/\psi$ are obviously OZI suppressed \cite{OZI},
so cannot be very large. One may write down the effective
lagrangians
$$L_1=g_1A_{1\mu}A_1^{\mu}\sigma, \;\;\;\;
{\mbox{ for}}\;\chi_{c1}-\sigma-\chi_{c1},$$ and
$$L_2=g_2\tilde F_{1\mu\nu} F_2^{\mu\nu}\sigma, \;\;\;\;
{\mbox{ for}}\;\chi_{c1}-\sigma-J/\psi,$$ where
$$\tilde F_{1\mu\nu}\equiv {1\over
2}\epsilon_{\mu\nu\alpha\beta}F_1^{\alpha\beta},$$ and
$A_{1\mu},\; A_{2\mu}$ correspond to axial vector $\chi_{c1}$ and
vector $J/\psi$ respectively.

It is supposed that the $\sigma$ exchange provides an attractive
potential $\propto {-e^{-m_{\sigma}r}\over r}$ between $\chi_{c1}$
and $\rho$ to construct a bound state. Apparently, the coupling is
OZI suppressed, and the binding is relatively loose.

More concretely, in Fig.1 (a), $\chi_c$ may convert into $J/\psi$
mainly via exchanging $\sigma$ particle with the constituent
$\rho$ meson. It is noted that $L_2$, which turns $\chi_{c1}$ into
$J/\psi$ is a p-wave interaction and proportional to the linear
momentum to guarantee the parity match. The differentiation may
result in an opposite sign to the potential between $\chi_{c1}$
and $\rho^0$ and  provide an effective repulsion.  Then the bound
state dissolves into free $J/\psi$ and $\rho$, and then a strong
decay of $\rho^0\rightarrow\pi^+\pi^-$ follows. Here, for a
general discussion, we ignore all the dynamical details and make
only an estimate on the order of magnitude. Since the branching
ratio of $\rho^0\rightarrow \pi^+\pi^-$ is almost 100\%, we can
suppose that the transition of the constituent of $Y(4260)$, i.e.
$\rho^0$ to $\pi^+\pi^-$ is overwhelming.

The total width is then,
\begin{eqnarray}
\Gamma(Y(4260)\rightarrow J/\psi\pi^+\pi^-)&=&{1\over 2M}\int {d^3
p_{_{J/\psi}}\over (2\pi)^3}{1\over 2E_{J/\psi}}{d^3
p_{_{\rho}}\over (2\pi)^3}{1\over
2E_{\rho}}(2\pi)^4\delta^4(M-P_{J/\psi}-P_{\rho})\cdot \nonumber
\\
&& |\mathcal{M}(Y(4260)\rightarrow J/\psi+\rho^0)|^2 \times
BR(\rho^0\rightarrow \pi^+\pi^-),
\end{eqnarray}
where $M$ is the mass of $Y(4260)$ and
$P_{J/\psi},p_{_{J/\psi}},P_{\rho},p_{_{\rho}}$ are the four- and
three-momenta of $J/\psi$ and $\rho$ respectively.

Comparing with Fig.1. (a), Fig.1 (b) involves an extra color
re-combination process which leads to a suppression, this
suppression factor is
\begin{eqnarray}
\frac{|\mathcal{M}(Y(4260)\rightarrow D\bar
D)|}{|\mathcal{M}(Y(4260)\rightarrow J/\psi\pi^+\pi^-)|}\propto
\alpha=\frac{1}{3}.
\end{eqnarray}
There may be a numerical factor $g$ coming from dynamics and it is
completely a non-perturbative QCD factor. For a rough estimate it
can be approximated as unity.

$Y(4260)\rightarrow J/\psi\pi^{+}\pi^{-}$ seems to be a three-body
decay, thus there could be a suppression from the phase space of
final states. However, in our picture of molecular state, it is
not a real three-body decay, instead, it is a two-step process,
namely first $Y(4260)$ dissolves into $J/\psi$ and $\rho^0$ and
then $\rho^0$ transits into $\pi^+\pi^-$. Since the  total width
is proportional to a two-body decay rate multiplied by the
branching ratio of $\rho^0\rightarrow \pi^+\pi^-$ which is 100\%
almost, there does not exist the phase space suppression factor at
all.

Due to the color re-matching factor, one can expect that the decay
rate of $Y(4260)\rightarrow J/\psi\pi^{+}\pi^{-}$ is about one
order larger than that of $Y(4260)\rightarrow D\bar{D}$. The
concrete dynamics may change this ratio more or less, but here we
just take this value from estimate of order of magnitude. This
value qualitatively coincides with the experimental results.

\section{More discussions and conclusion}

We suggest that the observed $Y(4260)$ is an s-wave molecular
state of $\chi_{c1}$ and $\rho^0$. It is natural to consider
another two partner molecular states, namely $\chi_{c0}+\rho^0$
and $\chi_{c2}+\rho^0$ in s-wave. Their spin-parity can be
different, but which one is dominant depends on the concrete
dynamics. For the simplest case, supposing they are also $1^{--}$,
we may expect that the molecular state of $\chi_{c2}+\rho^0$ is
only 40 MeV above 4260 MeV (supposing it has the same binding
energy as that for $\chi_{c1}+\rho^0$), on other side, the total
width of $Y(4260)$ is 90 MeV, thus this molecular state might be
hidden in the observed peak of $Y(4260)$, in other words, the
experimentally observed peak $Y(4260)$ may cover two close states.
Meanwhile, the molecular state of $\chi_{c0}+\rho^0$ could be 100
MeV below the central value of the peak and thus corresponds to a
new state which can be used as a test of the model. Namely, if
this partner resonance is observed in the future experiments, one
can claim that the molecular structure postulation may be correct,
otherwise, we need to consider other possible mechanisms to
suppress its production rate from dynamics or abandon the
molecular state interpretation. Moreover, the molecule of
$\chi_{c1}-\rho^0$ is a component of an isovector, so there may
exist another two components of the isotriplet, i.e.
$\chi_{c1}-\rho^{\pm}$ which may decay into $J/\psi
\pi^{\pm}\pi^0$ with comparable rates of $Y(4260)\to
J/\psi\pi^+\pi^-$. They may be experimentally observable. For the
molecular structure, $\sigma$ exchange between $\chi_{c1}$ and
$\rho^0$ may result in an attractive potential which binds them
into a molecule. Since the coupling is OZI suppressed, the binding
is relatively loose.

In our scenario, the favorable decay mode of $Y(4260)$ is
$Y(4260)\rightarrow J/\psi\pi^{+}\pi^{-}$. The molecular structure
of $Y(4260)$ results in different decay pattern from $\psi(3770)$
which is supposed to be a pure $c\bar c$ charmonium. Namely if we
take the $D\bar D$ mode as a standard, the rate of
$Y(4260)\rightarrow J/\psi\pi^{+}\pi^{-}$ is larger than that of
$Y(4260)\rightarrow D\bar D$ by an order.

By contraries, in the hybrid charmonium structure
\cite{Zhu,Kou,Close}, where a color-octet $c\bar c$ system is
bound with an octet valence gluon, since gluon is flavor-blind, it
has the same coupling to $q\bar q \;(q=u,d)$ and $s\bar s$, thus
besides a small suppression from the phase space,
$Y(4260)\rightarrow J/\psi\pi^{+}\pi^{-}$ and $Y(4260)\rightarrow
J/\psi K\bar K$ should be comparable unless there exist certain
mechanisms to suppress $K\bar K$ production. In the
diquark-anti-diquark picture of $[cs][\bar{c}\bar{s}]$, the mode
$Y(4260)\rightarrow J/\psi K\bar K$ overwhelms $Y(4260)\rightarrow
J/\psi\pi^{+}\pi^{-}$. In our picture of molecular state, the mode
of $Y(4260)\rightarrow J/\psi K\bar K$ can only be realized via
final state interaction $\pi^+\pi^-\rightarrow K\bar K$, so that
the rate of $Y(4260)\rightarrow J/\psi K\bar K$ is much smaller
than that of $Y(4260)\rightarrow J/\psi\pi^{+}\pi^{-}$.

Let us turn to a subtle and difficult subject, the production of
$Y(4260)$ in $e^+e^-$ collisions. The production may occur via the
so-called hairpin mechanism \cite{hairpin} which does not suffer
from the suppression due to color matching. It seems that it has a
larger production rate than the direct (non-resonant) production
of $D\bar D$ at first glimpse. However, a detailed analysis
indicates that unless the energy $\sqrt s$ of $e^+e^-$ collisions
can be precisely tuned to 4260 MeV, the energy conservation
demands production of other hadrons such as pions in company with
$Y(4260)$, and the constraint from the final product phase space
would greatly suppress its production rate. To achieve concrete
values one must carry out model-dependent calculations and it is
beyond the scope of this work.

One more observation is that $\rho^0$ only decays into
$\pi^+\pi^-$, but not $\pi^0\pi^0$, therefore, if the molecular
picture is right, the mode of $Y(4260)\rightarrow
J/\psi\pi^{0}\pi^{0}$ must be very suppressed. Moreover, since in
our picture $\pi^+\pi^-$ are produced from the real
$\rho^0-$meson, the measured invariant-mass spectrum of
$\pi^+\pi^-$ should peak up at $m_{\rho}$. Looking at the figure
(Fig. 3 of \cite{Babar}), the dipion mass distribution of
$Y(4260)\rightarrow J/\psi\pi^+\pi^-$ seems to show some fine
structures. Since the spectrum is due to $\rho^0\rightarrow
\pi^+\pi^-$ decay, it is a p-wave structure. The authors of
ref.\cite{Babar} indicate that the observed $\pi\pi$ spectrum is
somehow rather an s-wave comparing with the Mote-Carlo results,
thus the molecular interpretation offers a non-standard
interpretation for the bump. More precise experiments in the
future may give a decisive conclusion. Thus the possibility that
$Y(4260)$ is a molecular state, is indeed worth careful studies.

Now let us draw a brief conclusion. We propose that the newly
observed $Y(4260)$ is a molecular state of $\chi_{c1}$ and
$\rho^0$, and our analysis indicates that this picture
qualitatively coincides with the experimental data. We naturally
explain why the rate of $Y(4260)\rightarrow \pi^+\pi^-J/\psi$ is
larger than that of $Y(4260)\rightarrow D\bar D$, namely why
$Y(4260)$ is only observed in $e^+e^- \rightarrow ISR
\pi^+\pi^-J/\psi$, but not in $Y(4260)\rightarrow D\bar D$. We
have also made predictions on existence of two other components of
the isotriplet, $\chi_{c1}-\rho^{\pm}$ which may be observed in
channels $J/\psi\pi^{\pm}\pi^0$, and the extra partner resonances
$\chi_{c0(c2)}-\rho^{0}$ along with their isotriplet components.
It is suggested that the state of $\chi_{c0}+\rho^0$ may be
distinguished from $Y(4260)$ and can be experimentally measured,
so should serve as a test of the model. The future experiments
will collect more data and confirm or negate the various
theoretical models as well as ours. For such experiments besides
the B-factories, BES and CLEO are also ideal places.

In our model, we only discuss the qualitative characteristics and
make estimate of order of magnitude, but ignore all the dynamics.
Definitely all the details of dynamics may change the numbers
quite much, but we hope that the qualitative conclusion and
analysis would remain unchanged, because they are independent of
the dynamical details.

\vspace{0.6cm}

\noindent Acknowledgment:

This work is  supported by the National Natural Science Foundation
of China. We are grateful to Prof. K.T. Chao, Dr. S.W. Ye and S.L.
Zhu for helpful discussions.

\end{document}